\documentclass[11pt,fleqn]{article}

\usepackage[english]{babel}
\usepackage{amsmath, amssymb, multicol, amsfonts, amsthm, amscd, epsfig, enumerate, xcolor}
\usepackage[latin1]{inputenc}
\usepackage[T1]{fontenc}
\usepackage{graphicx}
\usepackage{geometry}

\newtheorem{theorem}{Theorem}

\newtheorem{corollary}[theorem]{Corollary}

\newtheorem{remark}[theorem]{Remark}

\begin{document}

\author{
\emph{\textbf{Alessandro Ramponi}}\\
Department of Economics and Finance, University of Rome Tor Vergata\\
Via Columbia 2, \ 00133 Rome, \ Italy\\
alessandro.ramponi@uniroma2.it
	\\
\emph{\textbf{M. Elisabetta Tessitore}}\\
Department of Economics and Finance, University of Rome Tor Vergata\\
Via Columbia 2, \ 00133 Rome, \ Italy\\
tessitore@economia.uniroma2.it }

\title{Educational programs and crime: a compartmental model approach}
\maketitle

\begin{abstract}
In this paper, we present a mathematical model to describe the temporal evolution of delinquent behavior, treating it as a socially transmitted phenomenon influenced by peer interactions, thus similar to an epidemic. We consider a compartmental framework involving three ordinary differential equations to describe the dynamics among the three population groups:  individuals not incarcerated (susceptible), incarcerated offenders, and incarcerated offenders participating in an educational program. Transitions between the groups are governed by interaction-based mechanisms that capture the influence of peer effects in the spread of criminal behavior. The model revealed three equilibrium states: a delinquence free equilibrium, an equilibrium where no criminals attend an educational program, and a coexistence equilibrium. The basic reproduction number, $R_0$, was derived, and a sensitivity analysis revealed the key parameters that influence the system's stability. The model thus provides a quantitative basis for evaluating the effectiveness of rehabilitation strategies in correctional settings. Numerical simulations and an empirical application illustrate the qualitative properties of the model and show how parameter variations influence system behavior.

\end{abstract}

\thispagestyle{empty}




\section{Introduction}

In Italy, education is a constitutionally guaranteed right and within prisons it becomes a fundamental treatment element for the re-socialisation and reintegration of the detained person into society. It should be pointed out from the outset that the number of detained persons accessing courses, and the courses themselves, has been steadily increasing over time. Prison education should have the same characteristics in terms of curricula and teaching methods as outside schools and provide, at least on paper, the possibility for student-offenders to follow a path from primary school to university.
The organisation of university courses in prison is considered a good Italian practice compared to other international contexts\footnote{see report Antigone https://www.rapportoantigone.it/ventesimo-rapporto-sulle-condizioni-di-detenzione/istruzione/.}, where it is not always guaranteed or provided for\footnote{ A prominent initiative between Rebibbia Prison and the University of Rome Tor Vergata is the project "Universit\'a in prigione", which provides inmates with access to higher education, enabling them to attend courses and earn degrees. The program relies on volunteer faculty participation: M. Elisabetta Tessitore teaches General Mathematics within this framework.}.

One unlawful method of destabilizing human civilization is through criminal activity. Addressing this issue is essential, as it has persisted for centuries \cite{SHIV}. 
Precisely defining crime is inherently challenging, as each society operates according to its own norms and values. Nevertheless, crime remains a major sociological concern and has been extensively studied in the scientific literature \cite{Pa}. In this context, researchers have developed increasingly sophisticated mathematical models, with particular attention to recidivism and the impact of released offenders on crime dynamics \cite{BL}, \cite{BBG}.
Blumstein \cite{BLU} reviewed the use of these models to determine cost-effective crime reduction strategies, such as optimizing police deployment and incarceration policies.

Over time, the focus shifted from macro-level crime trends to micro-level analyses of individual criminal careers. Researchers examined how individuals' crime participation evolved over time, from initiation to desistance \cite{BCR}, \cite{BLUC} and introduced the concept of selective incapacitation \cite{GAa}. This policy suggested incarcerating high-rate offenders would most effectively reduce crime but faced ethical criticism due to the difficulty of accurately predicting future criminal behavior. Later studies revealed that incarcerated populations had a much higher rate of offending than the general population, even without explicit selective incapacitation policies \cite{CCB}, hence there is a contagious effect. 

Mathematical models are increasingly being utilized to analyze criminal behavior and can complement traditional crime prevention strategies.
Besides this application, mathematical models have long been employed to understand the spread of infectious diseases and their dynamics \cite{KM}. While policymakers were already familiar with such models from past disease outbreaks, their significance and adaptability became especially evident during the COVID-19 pandemic, see e.g. \cite{BF2020}, \cite{CRS25}, \cite{CPW22} and \cite{RT23}. For the first time, policy decisions on a global scale were heavily influenced by modeling efforts. With nearly a century of development \cite{KM} and continuous advancements in computing power, compartmental models have evolved into powerful tools for analyzing and predicting disease transmission. Beyond forecasting, these models assist in shaping health policies by comparing intervention and relaxation strategies, estimating epidemiological metrics such as the basic reproduction number, and optimizing resource allocation.

The application of mathematical models extends beyond infectious diseases. Given the social nature of issues such as gang activity, youth delinquency, terrorism, and corruption \cite{G}, \cite{GAA}, these models are increasingly used to study such behaviors \cite{LT}, \cite{So}. Just as modeling informed prevention and treatment strategies during the COVID-19 pandemic, similar approaches hold potential for aiding crime control agencies. Various mathematical frameworks including ordinary and partial differential equations, agent-based modeling, game theory, and statistical physics have been utilized to study criminal behavior, see e.g. \cite{DP},  \cite{SB}, \cite{GSM}, \cite{McM}, \cite{Mis}, \cite{MBB}, \cite{CC}, \cite{MPS}, \cite{MS}, \cite{ABP}, \cite{RB}, \cite{SBB}, \cite{SDO}. Additionally, other mathematical approaches have been proposed \cite{SGA}, \cite{Tao}, \cite{YT}.

A primary motivation for employing compartmental models in criminological research lies in the analogy between the spread of epidemics and the diffusion of criminal behavior. Epidemics propagate through contact, and similarly, criminal tendencies are often transmitted via social interactions with delinquent peers \cite{G}, \cite{GAA}. This conceptual parallel has driven the development of dynamic models that treat criminal behavior as socially "contagious." The notion of crime as a socially transmitted phenomenon has been further explored in a number of recent studies \cite{McM}, \cite{Mis}, each focusing on different aspects of criminal dynamics. It has been widely suggested that individuals learn criminal behavior through interactions with offenders, and that increased exposure to criminal peers elevates the likelihood of adopting similar behaviors. For instance, the model in \cite{Mis} divides the population into four compartments, such as criminals and law enforcement, while neglecting individual differences in age, offense frequency, crime types, and criminal justice stages. Moreover, it postulates that the entry of new criminals into society is inversely related to the size of the police force.

In \cite{McM}, various compartmental structures are examined, with particular emphasis on a three-compartment system designed to study recidivism. These frameworks underscore the utility of compartmental models for capturing key mechanisms in the evolution of crime over time.

 A related body of work reinforces the suitability of such models by drawing explicit parallels between the spread of infectious diseases and the diffusion of criminal behavior \cite{BC}, \cite{KDM}. Criminal conduct, much like a virus, may propagate through networks of social contact. This perspective has informed compartmental analyses of specific forms of criminal activity, including gang violence, youth delinquency, terrorism, and corruption, the latter being especially consequential due to its role in exacerbating inequality, instability, and social unrest. 
 
 In \cite{SC}, a four compartments model is proposed with an intervention program in jail. It identifies three equilibrium states two of them are delinquence free, while the  coexistence equilibrium is treated numerically. In \cite{CM}  a three compartments model is proposed where the population is divided into non-offenders, offenders not in prison and incarcerated offenders. They study the stability of two equilibrium points.

In our paper, we contribute to this strand of literature by  providing a relatively simplified analysis of the population dynamics related to crime, incarceration and education. The model presented here assumes homogeneous populations. We do not account for variations in age, offense frequency, crime type, or actions taken at different stages of the criminal justice process (e.g., arrest, conviction, sentencing, parole). 
We aim to calculate and  explore the spread of crime and the dynamics of incarceration when an educational program is available in jail.

Using a compartmental epidemiological approach, the population is classified into three groups: susceptibles $X$ people who have no criminal behavior, and the infectious compartments  $E$  and $I$, who are criminals in jail attending an educational program and  criminals in prison who are not enrolled in any learning initiative, respectively. The resulting mathematical model consists of a nonlinear system of three  ordinary differential equations, analyzed to determine equilibrium points and their stability, including the threshold parameter $ R_0$, which indicates the potential for crime extinction and it will be crucial to establish the stability of the system's three equilibrium points: a delinquence free equilibrium DF, an education free equilibrium EF and a coexistence equilibrium CE. We are going to show that if the reproduction number  $R_0$ is below a certain range, then the delinquence free equilibrium is stable, else if $R_0$ belongs to a specific interval the education free equilibrium is stable and finally if  $ R_0$ is big enough, then the coexistence equilibrium is stable. Numerical simulations are conducted to validate the theoretical findings. Finally, an empirical application is presented, based on data provided by the Italian Ministry of Justice concerning prisoners' participation in vocational education courses. This example highlights both the promising features of the proposed model and the challenges associated with its empirical implementation.

The organization of the paper is the following. In Section 2, the mathematical model of crime is presented and its equilibria are evaluated. Section 3 studies the reproductive number $ R_0$ and  the stability of equilibria. Section 4 reports the results of numerical simulations and a case study. Finally, in Section 5, the conclusions are drawn and possible extensions are commented.

\section{The  model}

The population is categorized into two groups: offenders and non-offenders. An offender is defined as an individual who commits an unlawful act that warrants intervention from authorities, such as receiving a fine, performing community service, or facing a restriction of freedom. At any given moment, offenders may or may not be incarcerated. The latter scenario could arise if the individual has not yet been apprehended by law enforcement, is awaiting trial or legal decisions while remaining outside of jail, or has been convicted but is serving a sentence that does not involve imprisonment. Let $X(t)$ represent the number of non-offenders or of offenders not currently in prison, $I(t)$ denote the number of incarcerated offenders, and $E(t)$ indicate the number of  offenders in prison, attending an educational program, at any time $t \geq 0$. These categories, or compartments, encompass the individuals within the population under study.
A compartmental model outlines the flow of individuals between different categories, governed by both linear and nonlinear dependencies on the state variables. As referenced in the Introduction, it is assumed that offenders outside of prison impact non-offenders (i.e., susceptible individuals) through social interactions \cite{SC}. The dissemination of criminal behavior is inherently a complex and dynamic phenomenon. However, for the purposes of this paper, a simplification is adopted, suggesting that the sole pathway for someone to become criminally active is through direct interaction with another person already engaged in criminal activity (mass-action transmission).
The total population, denoted as $N$, is subdivided into mutually exclusive compartments of individuals $X, I$ and $E$.
\begin{itemize}
	\item In the $X$  compartment there are vulnerable persons who could potentially engage in offending behavior through social interaction, whether influenced by persuasion, coercion, or mimicking a behavior in response to a stimulus, among other factors. It is reasonable to presume that, for a certain period ranging from minutes to days the individual remains outside of prison, as they have not been apprehended. The 'force of infection' is encapsulated by a parameter $\alpha >0$.
	\item  
	In the $I$ compartment there are the convicted people who do not attend an educational program. After spending an average duration of $\frac{1}{\gamma _I}$  in prison, where $\gamma _I> 0$ represents the release rate, a convicted individual transitions back to the non-offender category $X$. While some individuals may re-engage in offending behavior following their time in prison, it is a reasonable approximation to assume that, for a certain period whether minutes, hours, or days infractions do not recur. This assumption helps to reduce the complexity of the mathematical model by minimizing the number of parameters.
	\item  Finally, in the $E$ compartment there are offenders in jail who attend an educational program and have
	an average incarceration period of $\frac{1}{\gamma _E}$, where $\gamma _E > 0$ denotes the rate of release. The parameters $0<\varphi \leq 1$ and $0<\beta \leq 1$ represent the fraction of community members who either engage or abstain from the implemented education programs. Smaller values of $\beta$ reflect a lower level of public participation in these initiatives, while values approaching one indicate strong overall compliance.
\end{itemize}

A coincise description of the compartments is summarized in Table \ref{compart}.
The development of the model relies on the  assumptions  presented below: 
\begin{itemize}
\item  a homogeneously-mixed population - that is to say,  all individuals in the population/community are assumed to have an equal probability of coming into contact with one another;

\item inclusion of demographic processes (e.g., migration, births, or deaths unrelated to the crime being modeled).
There is a parameter $\Lambda > 0$ that accounts for inflows, such as births and immigration, and a parameter $\mu > 0$ that represents outflows, including deaths and emigration. The average duration within the system is given by $\frac{1}{\mu}$, which approximately corresponds to the life expectancy in the studied society;

\item


transitions from the X (susceptible) compartment to the I (infected) compartment, and similarly for transitions involving compartments I and E, are governed by mass-action dynamics [34, p. 24].
Mass-action incidence implies that the rate of new "infections" (i.e., transitions into the I compartment) is proportional to the product of the number of susceptible individuals and those already in the infected state;


\item the transition from \( I \) to \( X \) is modeled as occurring at a rate proportional to the number of infected individuals, i.e., \( \gamma_I I \), where \( \gamma_I\) is the exit rate from compartment $I$. Similarly, the transition from \( E \) to \( X \) is assumed to occur at a rate \( \gamma_E E \), with \( \gamma_E \) denoting the corresponding exit rate from the educational program.

\end{itemize}

\begin{table}[t] 
	\small
	\centering
	\begin{tabular}{l|l}
		\hline
		$X$ & Individuals within the population who are vulnerable to engaging in delinquent behavior.
		 \\
		$I$ & incarcerated individuals who are not in an educational program \\
			$E$ & incarcerated  individuals who are attending an educational program \\
		\hline
	\end{tabular}
	\caption{Compartment explanation.} 
	\label{compart}
\end{table}

\begin{table}[h] 
	\small
	\centering
	\begin{tabular}{l|l}
		\hline
		$\Lambda $&  "recruitment" rate
		\\
		$\mu $ & death rate
		\\
		$\alpha $ &   rate of incarceration \\
		$\gamma_E$ & rate of incarcerated attending an educational program who get  out of prison\\
		$\gamma_I$ &  rate of incarcerated not attending an educational program who get  out of prison \\
		$\varphi $ & rate of people who are incarcerated and decide to start an educational program \\
		$\beta $& rate of people who are incarcerated and decide to quit an educational program
		\\
		\hline
		
	\end{tabular}
	\caption{Parameters explanation.} 
	\label{tab:param}
\end{table}

\begin{figure}[h]
	\begin{center}
		\includegraphics[width=14cm, height=8cm]{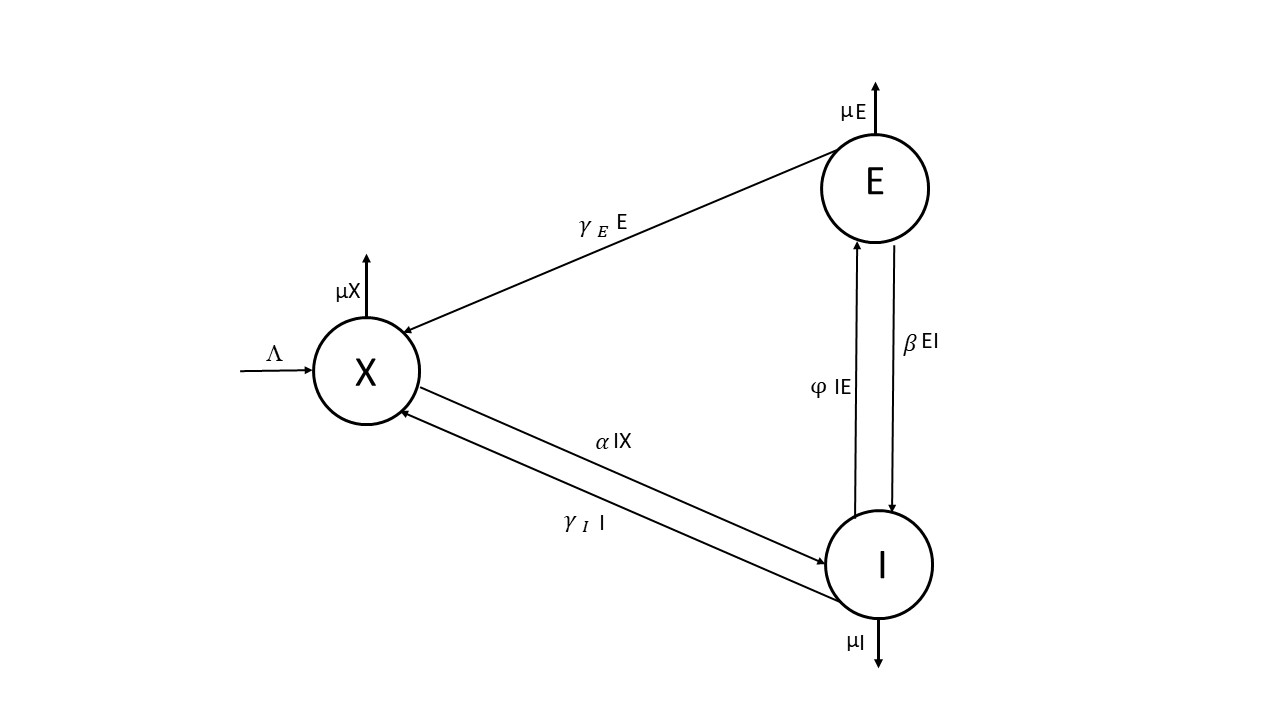}
	\end{center}
	
	\caption{The  model graph.}
	\label{grafo1}
\end{figure}



It is assumed that all the transition rates and parameters described in this study remain constant over time, as indicated in reference \cite{BC}, see Table \ref{tab:param} to recall the meaning of each parameter. This assumption simplifies the analysis by maintaining stability in these variables throughout the modeling process. The associated mathematical framework, built upon this assumption, is expressed through a system comprising three ordinary differential equations, which are designed to capture the dynamics of the system under consideration:

\begin{equation}
	\label{sistemanl}
	\left\{
	\begin{array}{ll}
		\displaystyle{\frac{dX}{dt}}(t)=\Lambda +  \gamma_E E(t)  + \gamma_I I(t)    -\alpha I(t)X(t)- \mu X(t)   & \ X(0)=x_0
		
		\\\\

		\displaystyle{\frac{dI}{dt}}(t) =   \alpha I(t)X(t)  +\beta  E(t) I(t) -\varphi E(t)I(t) - \gamma_I I(t)  -\mu I(t)  & \ I(0)= i_0
		\\ \\
		\displaystyle{\frac{dE}{dt}}(t) = \varphi E(t)I(t)-\beta E(t)I(t) - \gamma_E E(t)  -\mu E(t)  & \ E(0)=e_0,
		
	\end{array}
	\right.
\end{equation} 
where $x_0,e_0$ and $i_0$ are non--negative constants.
The first equation describes how individuals in the community $X$ are moving in and out of it. The term $\Lambda$ quantifies the flow entering in the susceptible compartment either moving in a particular dangerous area or by birth, while the parameter $\mu$ is the death rate that we assume constant in each compartment. The second  and the third equations describe the dynamics within the incarcerated population who can choose  to attend or not an educational program. Figure \ref{grafo1} shows the model graph.
Finally, $N(t)$ is the entire population, where
\begin{equation}
	\label{N}
	\left\{
	\begin{array}{ll}
		\displaystyle{\frac{dN}{dt}}(t)=\Lambda  - \mu N(t)   & \ N(0)=N_0=\frac{\Lambda}{ \mu}
		\\ \\
		N(t)=X(t)+E(t)+I(t).
		
	\end{array}
	\right.
\end{equation} 
Therefore we obtain $N(t)=\frac{\Lambda}{ \mu}, \quad \forall t\in [0, \infty)$.
It is assumed that the total population size equals $\frac{\Lambda}{\mu}$. While this approximation simplifies reality by assuming that inflow and outflow rates are roughly balanced, it allows for the study of population proportions in a two-dimensional framework and supports a manageable autonomous system.

\subsection{Equilibria }
Understanding equilibrium states is essential for analyzing the behavior of any dynamical system. A system in a stable equilibrium state resists small disturbances, meaning minor variations do not alter its long-term behavior. In contrast, if a system deviates from equilibrium following a small disturbance, the equilibrium is considered unstable.

Equilibrium states are determined by writing the isoclines and evaluating their intersections. In our model, three possible equilibrium states exist: the completely delinquence free equilibrium $DF$, the education free equilibrium $EF$, and the coexistence equilibrium, $CE$.

Let  $\rho:= \varphi - \beta$, we assume $\rho>0$  since  the treatment in the $E$ compartment is usually  better than in the $I$ compartment, where you can have permission to attend lectures or meet tutors, hence there should be fewer people moving from $E$ to $I$, then in the opposite direction.
We evaluate $X(t)=N_0-E(t)-I(t)$, we substitute it into (\ref{sistemanl}),   and we obtain:
\begin{equation}
	\label{sistemanEI}
	\left\{
	\begin{array}{ll}
		
		\displaystyle{	\frac{dI}{dt}(t) }=   I(t)[-(\rho+\alpha )E(t) -\alpha I(t)+\alpha  N_0-\gamma_I -\mu ] , & \ I(0)= i_0
		\\ \\
		\displaystyle{\frac{dE}{dt}(t)} = E(t)[\rho I(t) - \gamma_E-\mu ],& \ E(0)=e_0
		
	\end{array}
	\right.
\end{equation}

The equilibrium points are the intersecctions of the vertical   and the horizontal  isoclines in the $E-I$ plane.

We write the vertical isoclines $\displaystyle{\frac{dE}{dt}}(t) =0$:

$$
\frac{dE}{dt}(t) = E(t)[\rho I(t) - \gamma_E-\mu ]=0 ,
$$
hence they are: 
$E(t)=0 \mbox{ and }\displaystyle{I(t)=\frac {\gamma _E+\mu }{\rho }}.$ 

We write the horizontal isoclines $\displaystyle{\frac{dI}{dt}}(t) =0$:
$$
\frac{dI}{dt}(t) =   I(t)[-(\rho+\alpha )E(t) -\alpha I(t)+\alpha  N_0-\gamma_I -\mu ] =0
$$

hence they are: $I(t)=0$ and the straight line $ r:$ $I(t)=	-\frac {\rho +\alpha}{\alpha }E(t)+N_0\
-\frac {\gamma_I +\mu }{\alpha 
}.
$

Remarking that $r$  always has a negative slope and evaluating the intersections between the isoclines we obtain the following equilibrium points:
$$
P_1=(0,0), \qquad P_2=\left (0,N_0
-\frac {\gamma_I +\mu }{\alpha 
}\right )
\qquad \mbox{and}\qquad 
P_3=\left (-\frac {\alpha (\gamma _E+\mu )}{\rho (\rho + \alpha )}+
\frac {\alpha N_0-(\gamma _I+\mu )}{\rho + \alpha }
,\frac {\gamma _E+\mu }{\rho }\right )
$$
We are going to analyze three different cases.

\paragraph{Case 1. }
$$
N_0
-\frac {\gamma_I +\mu }{\alpha }>\frac {\gamma _E+\mu }{\rho }.$$

\begin{figure}[h]
	\begin{center}
		\includegraphics[width=14cm, height=8cm]{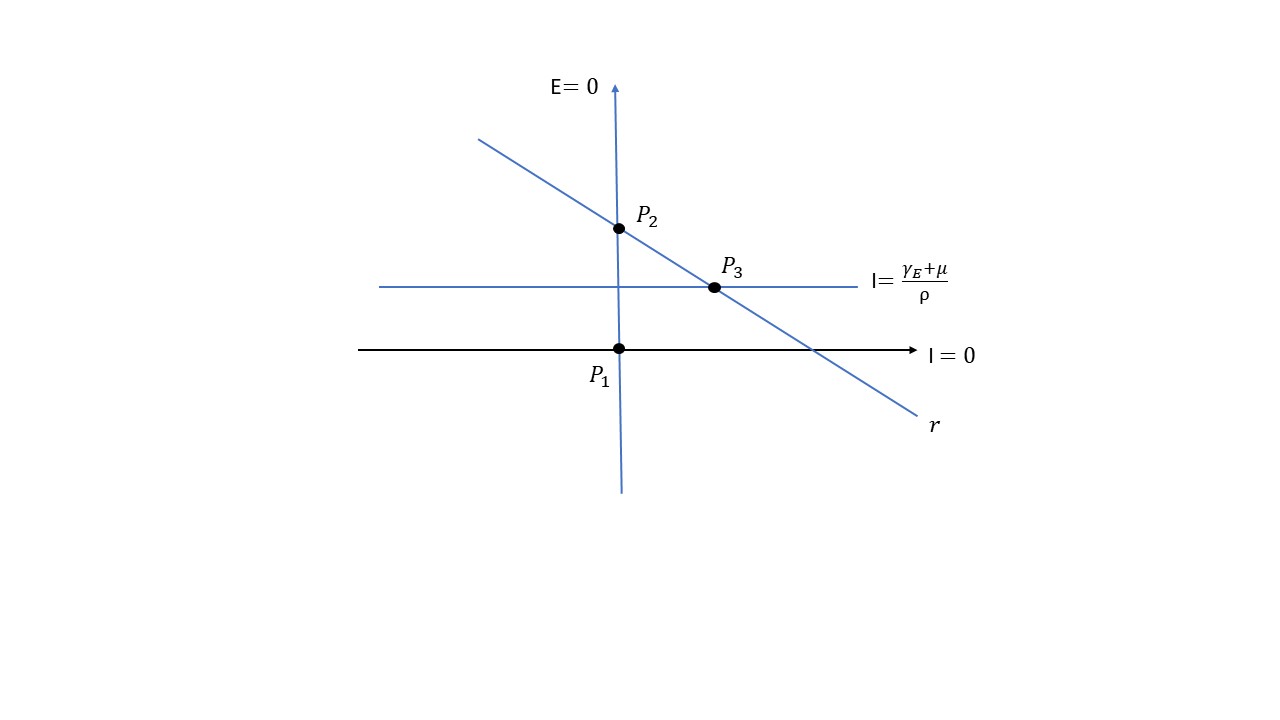}
	\end{center}
	
	\caption{Case 1}
	\label{C1}
\end{figure}

Beside the delinquence free equilibrium $P_1$, we have an education free equilibrium $P_2$ and an coexistence  equilibrium $P_3$ (see Figure \ref{C1}).
\paragraph{Case 2. }

$$
N_0
-\frac {\gamma_I +\mu }{\alpha }=\frac {\gamma _E+\mu }{\rho }.$$

\begin{figure}[h]
	\begin{center}
		\includegraphics[width=14cm, height=8cm]{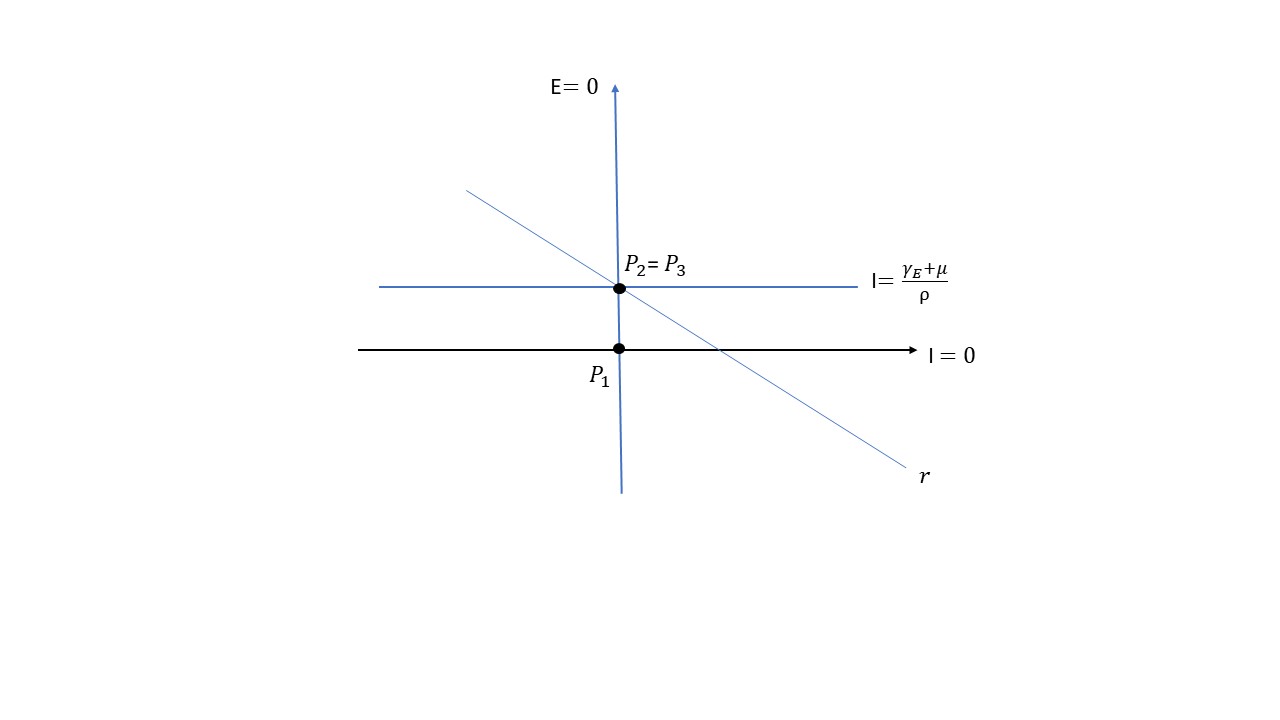}
	\end{center}
	
	\caption{Case 2}
	\label{C2}
\end{figure}
In this case, there is no coexistence equilibrium  (see Figure \ref{C2}).
\paragraph{Case 3. }

$$
N_0
-\frac {\gamma_I +\mu }{\alpha }<\frac {\gamma _E+\mu }{\rho }.$$

\begin{figure}[h]
	\begin{center}
		\includegraphics[width=14cm, height=8cm]{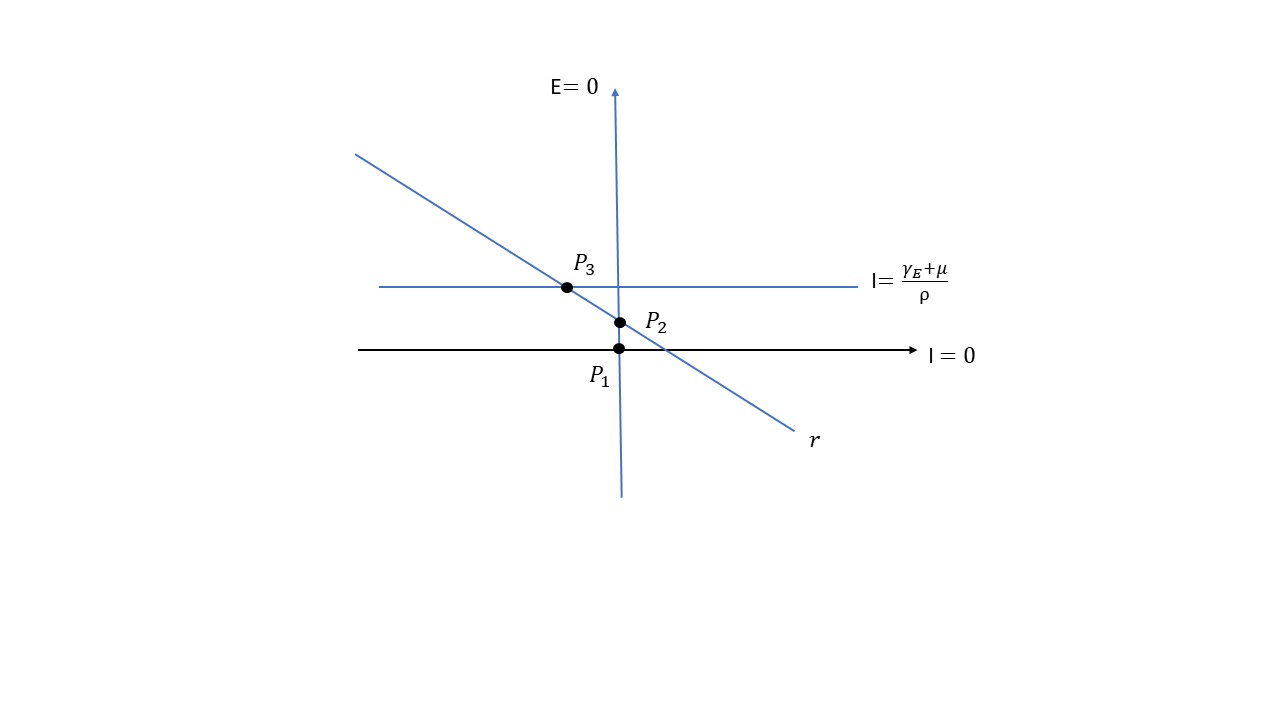}
	\end{center}
	
	\caption{Case 3}
	\label{C3}
\end{figure}
This last case presents an equilibrium with a negative $E$ component, hence it is not an interesting case  (see Figure \ref{C3}).

\medskip

We are interested in the first two cases, i.e. if
$$
N_0\geq \frac {\gamma_I +\mu }{\alpha 
}+
\frac {\gamma _E+\mu }{\rho }.
$$

\section{Calculation of the reproductive number ${R_0}$}

The persistence of a disease within a population depends on a critical threshold known as the basic reproductive number, $R_0$. This value represents the average number of individuals an infected person will transmit the disease to when entering a fully susceptible population. It reflects both the transmission dynamics of the disease and the contact patterns within the population.

If $ R_0>1$, an epidemic will occur, meaning the infection will persist within the population indefinitely unless behavioral changes or interventions are introduced. Conversely, if 
$R_0<1$, the disease will gradually disappear. Because $R_0$
determines the level of intervention needed to prevent an epidemic, it is a crucial parameter for public health planning. The primary objective is to implement mitigation strategies that reduce 
below the threshold of $1$.

The next-generation operator method is used to calculate $ R_0$. In this approach, compartments $E$, and $I$ represent the infective stages, while $X$ denotes the non-infective compartment. 

The right-hand side of the infected compartments is then decomposed into the form $ {F_1}$ and $ {V_1}$ as delineated below:

\begin{equation}
	\label{matriceF1s}
	 {F_1}=
	\begin{pmatrix}
		\alpha XI 
		\\
		0
	\end{pmatrix}
\end{equation}
and
\begin{equation}
	\label{matriceV1s}
 {V_1}=
	\begin{pmatrix}
		
		\rho EI+\gamma _I I  + \mu I
		\\\\
		-\rho EI+\gamma _E E  + \mu E 
	\end{pmatrix}
\end{equation}
Through this decomposition, two matrices, ${F}$ and $ {V}$,  with $x_j$ representing the infective compartments, are derived and they are evaluated at $\left (\frac{\Lambda}{\mu},0,0\right )$:
\begin{equation}
	\label{matriceFs}
	 {F}=\left [\frac{d{F_1}}{dx_j}\right ]=
	\begin{pmatrix}
		\alpha \frac{\Lambda}{\mu}&0
		\\
		0&0 
	\end{pmatrix}
\end{equation}
and 
\begin{equation}
	\label{matriceVs}
	{V}=\left [\frac{d{V_1}}{dx_j}\right ]=
	\begin{pmatrix}
		\gamma_I  +\mu &0
		\\
		0&\gamma_E +\mu 
	\end{pmatrix}
\end{equation}

The next--generation matrix ${K}$ is defined as  ${K}= {F}\; {V}^{-1}$ therefore

\begin{equation}
	\label{matriceKs}
	 {K}=
	\begin{pmatrix}
		\frac{\alpha \Lambda}{\mu (\gamma_I +\mu)}&0
		\\
		0&0
	\end{pmatrix}
\end{equation}

which has two eigenvalues: $0$ and ${R_0}=\frac{\alpha \Lambda}{\mu (\gamma_I +\mu)}$, hence $ {R_0}$, the spectral radius of $ {K}$, is:

\begin{equation}
	\label{R0s}
	 {R_0}=\frac{\alpha \Lambda}{\mu (\gamma_I +\mu)}.
\end{equation}

\begin{remark}
	Focus the attention on:
	\begin{itemize}
		
	\item the value $R_0$ depends on the parameters $\alpha$ and $\gamma_I$ (other than the demographic - hence exogenous - parameters $\Lambda$ and $\mu$). Therefore,  while $R_0$ is the quantity that determines the asymptotic behavior of the system, it does not depend on the educational parameters. Any educational strategy does not change directly the value of $R_0$;
		\item 
 the elasticity, which  measures the responsiveness of $ {R_0}$ to changes in $\alpha $ and $\gamma _I $.
	$$
	\nu _{\alpha ,{R_0} }=\frac{\partial 	{R_0}}{\partial 	\alpha}\cdot \frac{\alpha }{ {R_0}}=1
	$$
	and
		$$
		\nu _{\gamma _I , {R_0} }=\frac{\partial 	 {R_0}}{\partial 		\gamma_I}\cdot \frac{\gamma_I  }{ {R_0}}=-\frac{\gamma_I}{\gamma_I +\mu}.
	$$
	The elasticity $\nu _{\alpha ,{R_0} }=1$ means that the percentage change both for ${R_0}$ and  the conctact rate $\alpha $ is the same, 
	while the elasticity $\nu _{\gamma _I ,{R_0} }<0$   means that $ {R_0}$ decreases when $\gamma_I$ increases, and vice versa. In other words, the elasticity of ${R_0}$ is directly proportional to the average time $\frac{1}{\gamma_I}$ spent in prison by the incarcerated not attending any educational program.
	\end{itemize}
\end{remark}
	
\subsection{Stability of the equilibria}
The system has three equilibrium states: two of them  where  delinquents not performing an educational program are absent and one where all populations coexist. But what determines which equilibrium state is observed? This depends on the stability of the system at each equilibrium.

To assess stability, we analyze the Jacobian matrix of the nonlinear system. According to linear stability analysis, an equilibrium state is considered stable if all eigenvalues of the Jacobian have negative real parts. This approach helps identify the conditions required for each equilibrium state to remain stable. In what follows, we omit the time dependence in $X,I,E$.

The Jacobian matrix $ J$ for the right--hand side of system (\ref{sistemanl}) is:
{\small
	\begin{equation}
		\label{matricejs}
			J=
		\begin{pmatrix}
			- \alpha I  - \mu&
			\gamma_I -\alpha X&
			\gamma_E 
			\\\\
			\alpha I&
			\alpha X - \rho E- \gamma_I  -\mu &
			- \rho I 
			\\\\
			0 &
			\rho E&
			\rho I- \gamma_E   -\mu
		\end{pmatrix}.
	\end{equation}
}

\begin{theorem} \label{mainTheorem}
	The delinquence free equilibrium DF, 	
	
	$$
	\displaystyle{{DF}=({X^*},{I^*},{E^*})=\left (\frac{\Lambda}{\mu},0,0\right )
	}
	$$
	
	is locally asymptotically stable if $ {R_0} < 1$.
	
	The education free equilibrium EF, 	
	
	$$
	\displaystyle{{EF}=({X^*},{I^*}, {E^*})=\left (  \frac{\gamma_I + \mu}{\alpha}, N_0-\frac{\gamma_I + \mu}{\alpha}, 0\right )
	}
	$$
	is asymptotically locally stable if
	$$
	1<{R_0} < \overline C
	$$
where 
$$\overline C=	1+
	\frac{\alpha(\gamma_E+\mu)}{\rho (\gamma_I+\mu)}.
	$$

	The coexistence equilibrium, 	
	
	$$
	\displaystyle{{CE}=({X^*},{I^*},{E^*})=\left (\frac{\gamma_I - \gamma_E + N_0 \rho}{\alpha + \rho},\frac{\gamma_E + \mu}{\rho},-\frac{\alpha (\gamma_E +  \mu) + \rho(\gamma_I  + \mu  - N_0 \alpha )}{\rho (\alpha + \rho)}\right )
	}
	$$
	is asymptotically locally stable if
	$$
	 {R_0} >\overline C.
	$$
\end{theorem}

\begin{proof} 
	We are going to evaluate the Jacobian matrix (\ref{matricejs}) and its eigenvalues at the three equilibrium points.
	{\small
		\begin{equation}
			\label{matriceJpfree}
				J|_{{DF}}=
			\begin{pmatrix}
				- \mu&
				\gamma_I	-\alpha \frac{\Lambda}{\mu}&
				\gamma_E 
				\\\\
				0&
				\alpha \frac{\Lambda}{\mu} - \gamma_I  -\mu &
				0
				\\\\
				0 &
				0&
				- \gamma_E   -\mu
			\end{pmatrix}
		\end{equation}
	}
	whose eigenvalues are 
	$$
	\lambda _1=- \mu ,
	\qquad
	\lambda _2=		\alpha \frac{\Lambda}{\mu} - \gamma_I  -\mu
	=( {R_0}-1)(\gamma_I  +\mu), \qquad
	\lambda _3=- \gamma_E  -\mu.
	$$ 
	
	Therefore,  if ${R_0}<1$ then $ {DF}$ is locally stable. 
	
	$$
	J|_{EF} = \begin{pmatrix}       
		\gamma_I - N_0 \alpha & -\mu &  \gamma_E \\
		N_0 \alpha - \mu - \gamma_I &   0 &      -\rho (N_0 - \frac{\gamma_I + \mu}{\alpha}) \\
		0 &   0 &  \rho (N_0 - \frac{\gamma_I + \mu}{\alpha})- \mu - \gamma_E 
	\end{pmatrix} 
	$$
	
	whose eigenvalues are:
	
	$$
	\begin{array}{lll}       
		\lambda_1 & = & -\mu  
		\\\\
		\lambda_2 & = & -N_0 \alpha + \mu + \gamma_I =
		(1-{R_0})(\mu + \gamma_I )
		\\\\
		\lambda_3 & = & \rho (N_0 - \frac{\gamma_I + \mu}{\alpha}) - \mu - \gamma_E =
		\frac {\rho}{\alpha}({R_0}-1)(\mu + \gamma_I )
		- \mu - \gamma_E
	\end{array} 
	$$
	Therefore, if $ {R_0} > 1$, $\lambda_2$ is negative and $\lambda_3$ is negative if $ {R_0} < 1+
	\frac{\alpha(\gamma_E+\mu)}{\rho (\gamma_I+\mu)}=\overline C$,  hence if
	$$
	1< {R_0} < \overline C.
	$$
	${EF}$ is locally stable.

	$$
	 J|_{{CE}} = \begin{pmatrix}
		-\mu -\frac{\alpha \,\left(\mathrm{\gamma_E}+\mu \right)}{\rho } & \mathrm{\gamma_I}-\frac{\alpha \,\left(\mathrm{\gamma_I}-\mathrm{\gamma_E}+N_{0}\,\rho \right)}{\alpha +\rho } & \mathrm{\gamma_E}\\ \frac{\alpha \,\left(\mathrm{\gamma_E}+\mu \right)}{\rho } & 0 & -\mathrm{\gamma_E}-\mu \\ 0 & -\frac{\alpha \,\mathrm{\gamma_E}+\alpha \,\mu +\mathrm{\gamma_I}\,\rho +\mu \,\rho -N_{0}\,\alpha \,\rho }{\alpha +\rho } & 0 
	\end{pmatrix}
	$$
	
	Its eigenvalues are:
	$$
	\begin{array}{l} 
		\lambda_1=-\mu 
		\\ \\ 
		\lambda_2=-\frac{\alpha ( \gamma_E+\mu) -\sqrt{\left(\mathrm{\gamma_E}+\mu \right)\,\left(\alpha ^2\,\mathrm{\gamma_E}+\alpha ^2\,\mu +4\,\mathrm{\gamma_I}\,\rho ^2+4\,\mu \,\rho ^2+4\,\alpha \,\mathrm{\gamma_E}\,\rho +4\,\alpha \,\mu \,\rho -4\,N_{0}\,\alpha \,\rho ^2\right)}}{2\,\rho }
		\\ \\
		\lambda_3=-\frac{\alpha ( \gamma_E+\mu) +\sqrt{\left(\mathrm{\gamma_E}+\mu \right)\,\left(\alpha ^2\,\mathrm{\gamma_E}+\alpha ^2\,\mu +4\,\mathrm{\gamma_I}\,\rho ^2+4\,\mu \,\rho ^2+4\,\alpha \,\mathrm{\gamma_E}\,\rho +4\,\alpha \,\mu \,\rho -4\,N_{0}\,\alpha \,\rho ^2\right)}}{2\,\rho }
	\end{array}
	$$
	
	Since $N_{0}\alpha = {R_0}(\gamma_I+\mu) $, we derive
	$$
(\gamma_E+\mu )(\alpha ^2\gamma_E+\alpha ^2\mu +4\gamma_I\rho ^2+4\mu \rho ^2+4\alpha \gamma_E,\rho +4\alpha \mu \rho -4N_{0}\,\alpha \rho ^2)=
$$
	$$
	=\alpha ^2(\gamma_E+\mu)^2+
	(\gamma_E+\mu)4\rho ^2
	(  \gamma_I+\mu -{R_0}(\gamma_I+\mu ))+4\rho \alpha(\gamma_E+\mu)^2 =
	$$
	$$
	=\alpha ^2(\gamma_E+\mu)^2+4\rho ^2
	(\gamma_E+\mu)
	(  \gamma_I+\mu)(1 -{R_0})+4\rho \alpha(\gamma_E+\mu)^2
	$$
	Hence, if 	
	$$
	4\rho ^2
	(\gamma_E+\mu)
	(  \gamma_I+\mu)(1 -{R_0})+4\rho \alpha(\gamma_E+\mu)^2<0
	$$
	which can be rewritten as
	$$
	 {R_0} > 1+
	\frac{\alpha(\gamma_E+\mu)}{\rho (\gamma_I+\mu)}=\overline C,
	$$
	$\lambda _2$ and $\lambda _3$ are either negative or have  negative real part, therefore  ${CE}$ is locally stable. 
\end{proof}

\begin{remark}\label{remr0}
Our primary focus is to examine the shift from the EF to the CE state of balance: this occurs when $R_0$ reaches and surpasses the threshold of $\overline C$.
We consider $R_0$ fixed and study the conditions
$$
1 < R_0 < \overline C \ \ \ \mbox{and} \ \ \ \ R_0 > \overline C
$$ 
with respect to $\gamma_E$ and $\rho$. 

Assuming $R_0>1$, we notice that $R_0 = N_0 \frac{\alpha}{\gamma_I+\mu}$, from which we immediately get that
$$
R_0 \lessgtr \overline C \Longleftrightarrow R_0 \lessgtr 1 + \frac{R_0}{N_0} \frac{\gamma_E+\mu}{\rho},
$$  
implying
\begin{equation} \label{R0cond}
	R_0 \lessgtr \frac{N_0 \rho}{N_0 \rho - (\gamma_E+\mu)}.
\end{equation}
By considering fixed $R_0>1$, we may solve (\ref{R0cond}) in the $(\gamma_E,\rho)$-plane, thus obtaining
\begin{equation}
	\rho \lessgtr (\gamma_E+\mu) \frac{R_0}{N_0(R_0-1)}.
\end{equation}

	Therefore, we notice
	
\begin{itemize}
	
	\item 	if $\rho \equiv 0$ (i.e. $\beta \equiv \varphi$) then the coexistence equilibium doesn't exist; 
	\item the parameters which drive the flows between the compartments involving $E$ are $\gamma_E$ and $\rho$. These parameters determine the coexistence equilibrium value ${CE}$, and the critical threcshold  $\overline C$.
	\item the system will converge to the DF equilibrium if the condition $R_0 <1$ is satisfied: this can happen for sufficiently small $\alpha$ (the contact rate) and/or sufficiently large $\gamma_I$ (the inverse of mean time of permanence in the I compartment);
	
	\item CE equilibrium will approach the DF equilibrium as $\rho\to \infty $ meaning, that if everybody would attend an educational course, crime could disapear.
\end{itemize}
\end{remark}

\begin{corollary}
	\label{complex}
If 
	$$
{R_0} > 1+
\frac{\alpha(\gamma_E+\mu)}{\rho (\gamma_I+\mu)}	\frac{4\rho+\alpha}{4\rho }=1+\frac{\alpha(\gamma_E+\mu)}{\rho (\gamma_I+\mu)}(1+\frac{\alpha}{4\rho }),
	$$
then the Jacobian matrix evaluated at the coexistence equilibrium, has two complex conjugate eigenvalues. Moreover, their real part is
$$
\Re({\lambda_i}) = - \frac{\alpha (\gamma_E+\mu)}{2 \rho}, \ \ \ \ i=2,3.
$$ 
\end{corollary}
\begin{proof} 
Proceeding as in  Theorem \ref{mainTheorem}, we notice that the radicand term of the eigenvalues $\lambda _2$ and $\lambda _3$ is
	$$
\alpha ^2(\gamma_E+\mu)^2+4\rho ^2
(\gamma_E+\mu)
(  \gamma_I+\mu)(1 - {R_0})+4\rho \alpha(\gamma_E+\mu)^2=
	$$
	$$
	=(\gamma_E+\mu)[\alpha ^2(\gamma_E+\mu)+4\rho ^2
	(  \gamma_I+\mu)(1 - {R_0})+4\rho \alpha(\gamma_E+\mu)]
	$$
	 if 
	$$
	\alpha ^2(\gamma_E+\mu)+4\rho ^2
	(  \gamma_I+\mu)(1 - {R_0})+4\rho \alpha(\gamma_E+\mu)<0,
	$$
	 the conclusion follows. 

\end{proof}

\begin{remark}

To notice:
\begin{itemize}
	\item we have three types of equilibria dependent on $R_0$: delinquence free, education free and coexistence equilibria. 
If $R_0<1$ we have DF, while if $R_0>1$ we have the other two. Acting on $\alpha$ and $\gamma_I$ (as well as demographic parameters) we can have  $R_0<1$ or $R_0>1$;
	\item if $R_0>1$ (and therefore there are inmates) by modifying $\rho$ we can achive one of the two equilibria EF or CE, modifying the threshold level $\overline C$. For $\rho$ large enough, the threshold level is very close to 1 thus $R_0>\overline C$ and CE is the equilibrium of the system.

\end{itemize}

\end{remark}

\section{Numerical results}

This section presents the numerical results and is divided into two parts. First, we explore the behavior of the model through simulations with fixed parameter values, chosen to illustrate the theoretical properties established in Theorem \ref{mainTheorem}. In the second part, we conduct a case study using empirical data provided by the Italian Ministry of Justice, which requires a parameter estimation step in order to calibrate the model to real-world observations.

\subsection{Simulated examples}

We present a set of numerical simulations of the system (\ref{sistemanl}) to highlight its properties, as obtained in the theoretical section. In particular, we consider a baseline model specified by the following set of parameters: $\alpha = 1.0\times 10^{-6},  \ \ \gamma_E= 1, \ \ \beta = 1.0\times 10^{-4}, \ \  \phi = 1.2\times 10^{-4} $. Furthermore, we set $N_0 = 10^{6}$, $\mu = 0.01$ and $\Lambda = N_0\times \mu$, this last condition to ensure a constant population over time. The time zero values for the $I$ and $E$ compartments were set to $E_0 =N_0 \times 10^{-3}$ and $I_0=N_0 \times 10^{-2}$, respectively. Moreover, we assume the year as the time unit and consider $T = 100$.  All the numerical simulations were developed in MatLab R2024b, by using the \texttt{ode45} solver, which is based on an explicit Runge-Kutta formula.

\medskip

\begin{figure}[h]
	\begin{center}
		\includegraphics[width=16cm, height=10cm]{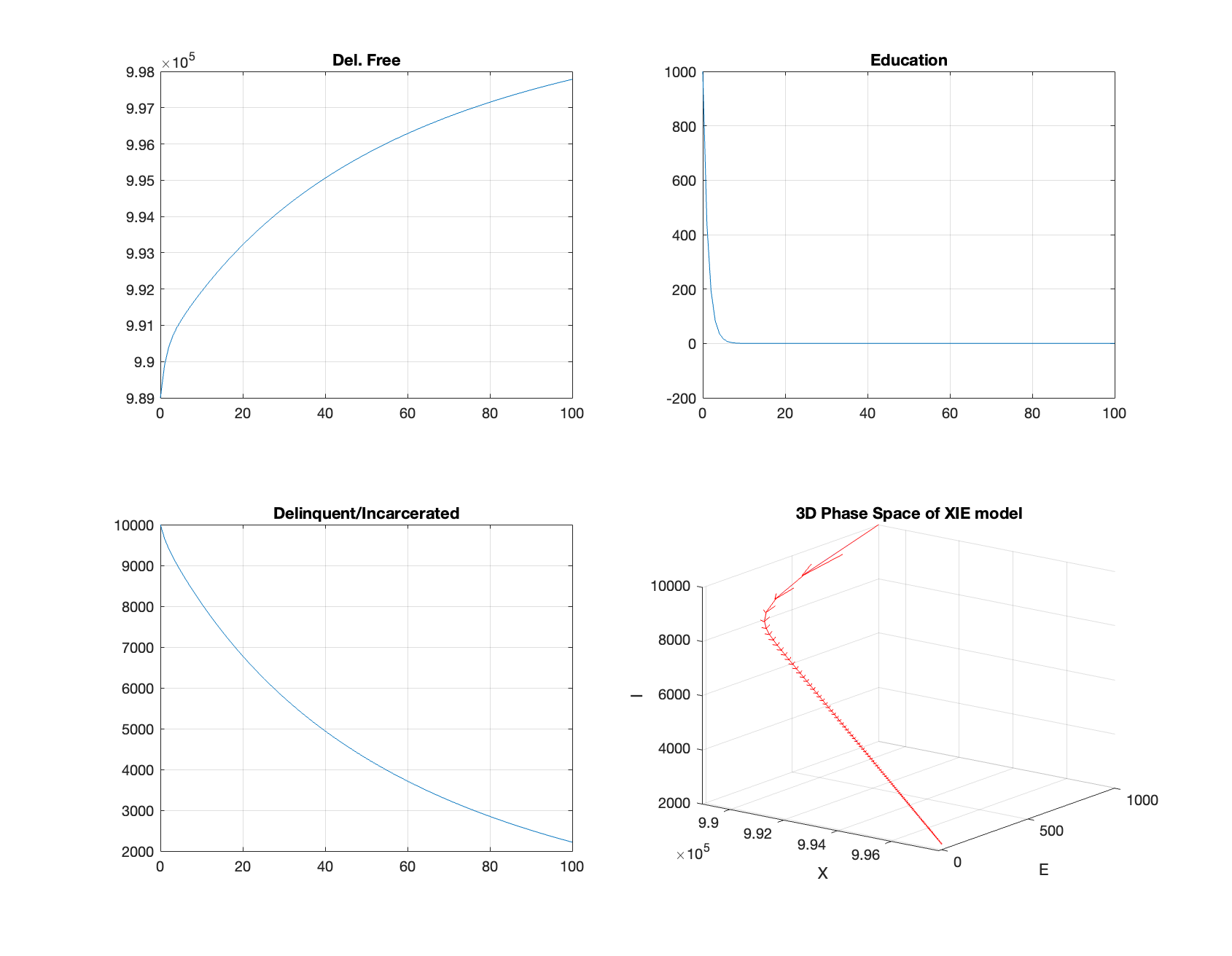}
	\end{center}
	
	\caption{Delinquence free equilibrium. Here $R_0 =0.99$. }
	\label{FigDF}
\end{figure}

\begin{figure}[h]
	\begin{center}
		\includegraphics[width=16cm, height=10cm]{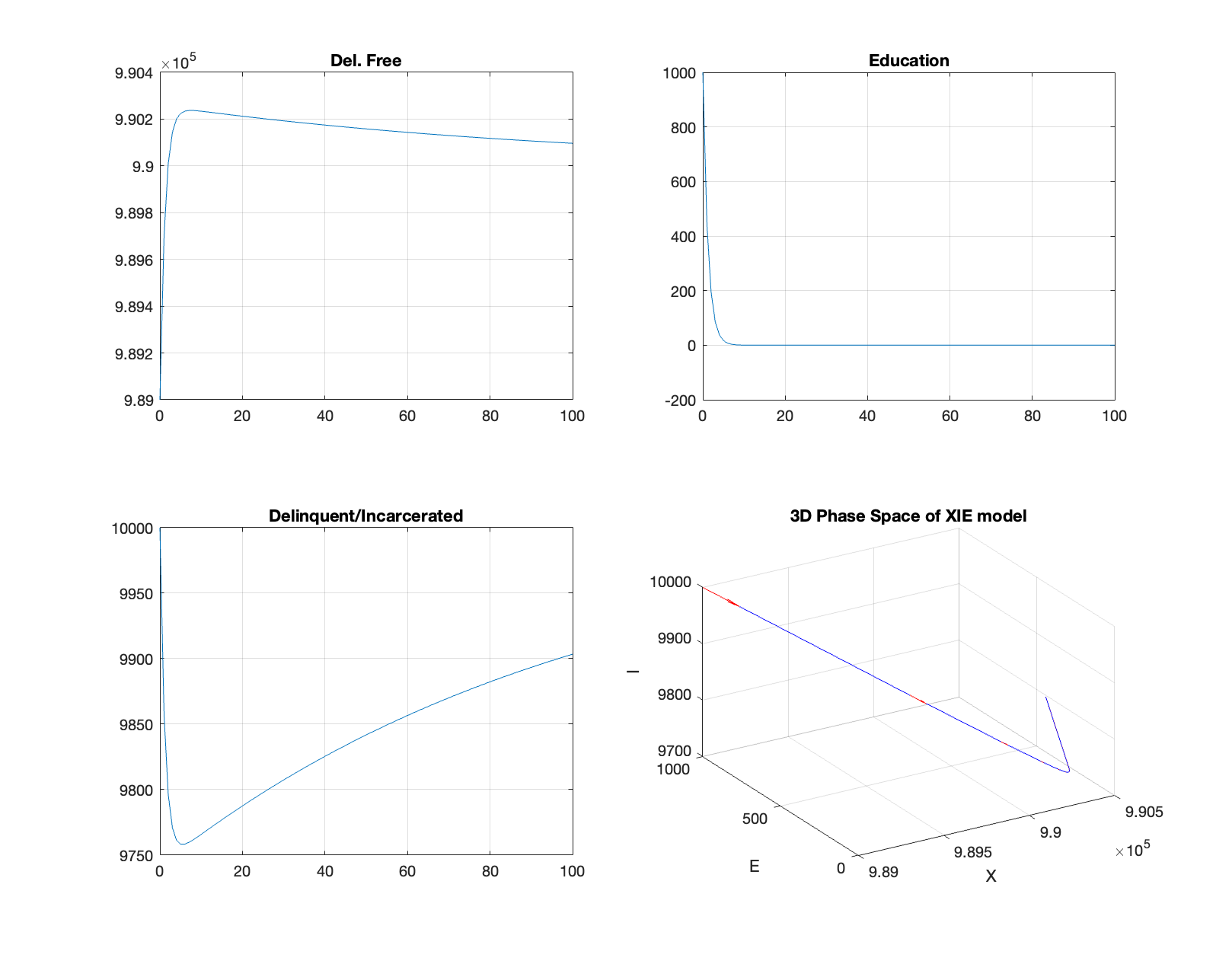}
	\end{center}
	
	\caption{Education-free equilibrium. Here $R_0 =1.01$, $\gamma_I = 0.98$.}
	\label{FigEF}
\end{figure}

\begin{figure}[h]
	\begin{center}
		\includegraphics[width=16cm, height=10cm]{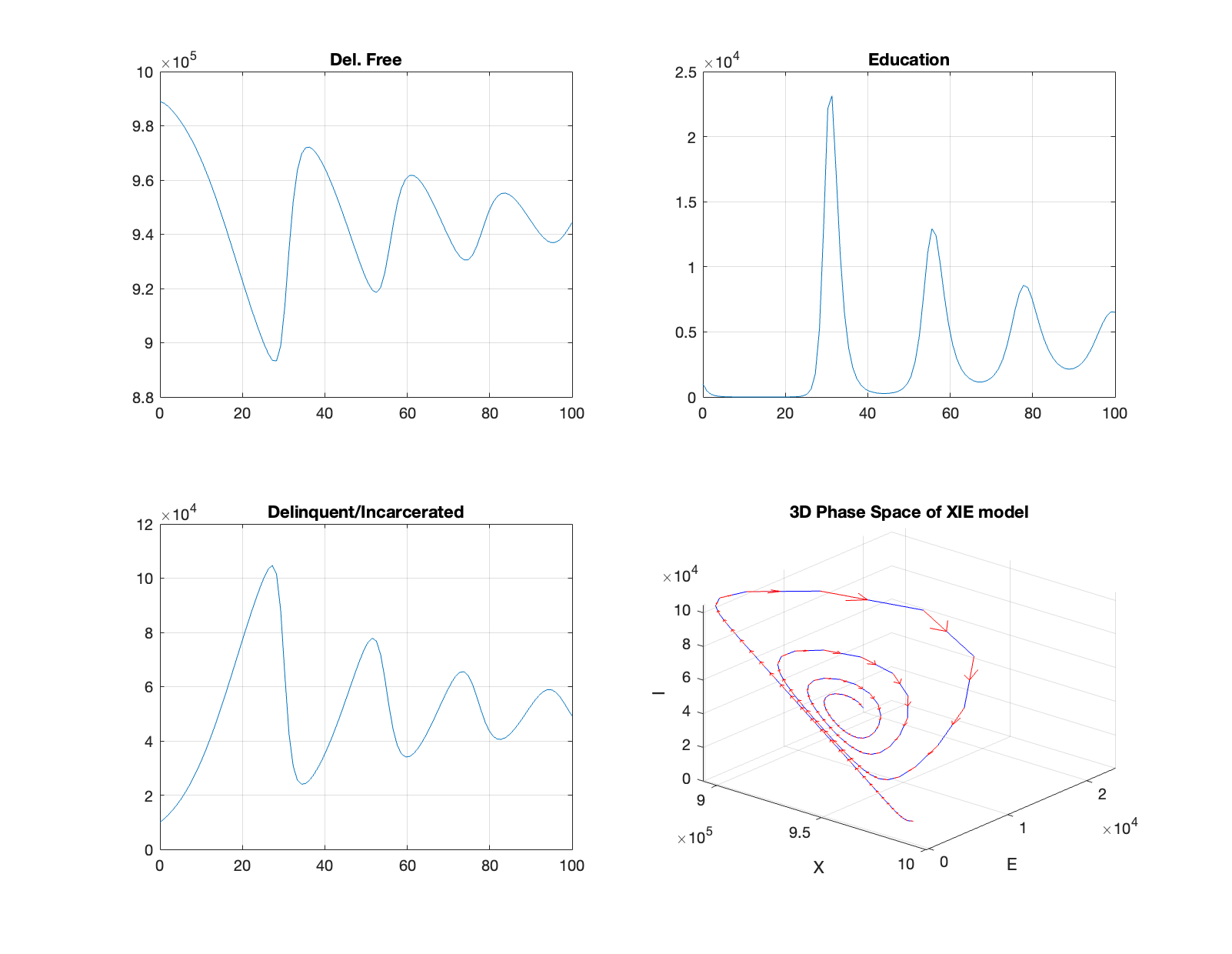}
	\end{center}
	
	\caption{Coexistence equilibrium. Here $R_0 =1.16$, $\gamma_I = 0.85$. }
	\label{FigCE}
\end{figure}

In the first set of experiment, we analyze  the system solutions in the three different ''regimes'', identified by the three kind of equilibria (DF, CF, and CE), corresponding to the conditions set by the value $R_0$, see Theorem \ref{mainTheorem}. These values were obtained by varying the parameter $\gamma_I $. Specifically, under the parameter choices described above, we identify three distinct scenarios, for which the corresponding time evolutions are illustrated in phase space:
\begin{description}
\item[$R_0 < 1$]: this case corresponds to the choice $\gamma_I =0.995 $. The DF equilibrium is reached quite quickly, see Figure \ref{FigDF}. This scenario seems largely unrealistic, as the interaction between the susceptible population and individuals prone to delinquency does not lead to a sustained increase in the prison population. As a result, the system evolves toward a state in which incarceration is entirely absent.
\item[$1 < R_0 < 1.05$]: in this case we have the EF equilibrium, obtained by setting $\gamma_I = 0.985$. The Figure \ref{FigEF} illustrates that the system evolves toward a state in which educational programs fail to take root within the prison population. The interaction dynamics are insufficient to sustain or expand the number of individuals participating in educational courses, leading to their eventual disappearance from the system.
\item[$R_0 > 1.05$]: the coexistence equilibrium is obtained by setting $\gamma_I = 0.850$.  The Figure \ref{FigCE} displays the time evolution of each compartment. In this scenario, the presence of complex eigenvalues gives rise to oscillatory dynamics (see Corollary \ref{complex}), which gradually dampen over time, ultimately converging to a stable coexistence equilibrium where all compartments persist.  This represents the most realistic scenario, as it reflects a system in which all three compartments remain non-empty in the long run.

\end{description}

In the second set of experiments, we want to analyze more deeply the effect of introducing an educational compartment. In particular, we would like to assess the impact of the parameters $\gamma_E$ and $\rho = \varphi-\beta$, which determine the in-out flows of the compartment E, on the value $R_0$ and $\overline C=1+\frac{\alpha(\gamma_E+\mu)}{\rho (\gamma_I+\mu)}$.

We set the parameters $\alpha$, $\gamma_I$, $\mu$, $\Lambda$ to the baseline model values,  and then we vary $\gamma_E$ and $\rho = \varphi-\beta$. Our main interest is to analyze the transition from the EF to the CE equilibrium: this happens when $R_0$ "cross" the value $\overline C$, see Remark  \ref{remr0}.

\begin{figure}[h]
	\begin{center}
		\includegraphics[width=12cm, height=8cm]{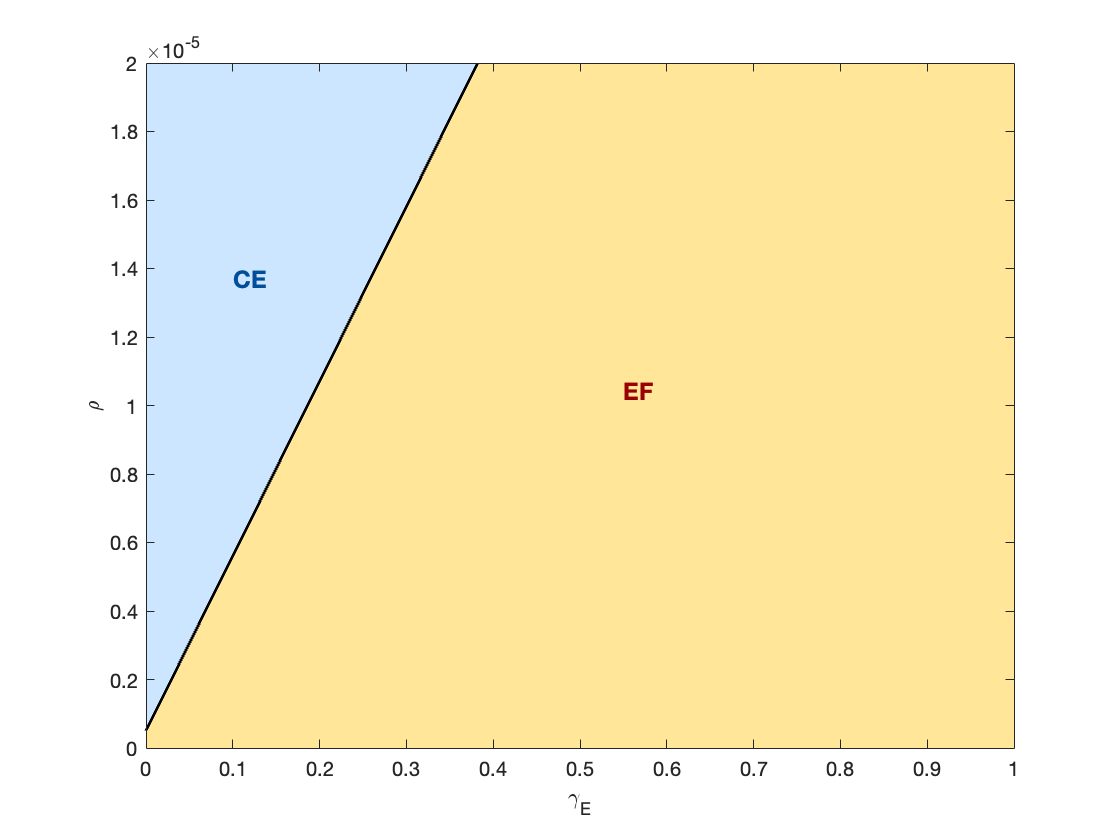}
	\end{center}
	\caption{Values of the parameters $\gamma_E$ and $\rho$ which determine the education free equilibrium, EF, and the coexistence equlibrium, CE. In this example, $R_0=1.02$.}
	\label{Fig_regions}
\end{figure}

Figure (\ref{Fig_regions}) shows the values of $\gamma_E$ and $\rho$ that determine the two types of equilibrium, education free EF and coexistence equilibrium CE.

\subsection{A real data example}

\begin{figure}[h]
	\begin{center}
		\includegraphics[width=12cm, height=8cm]{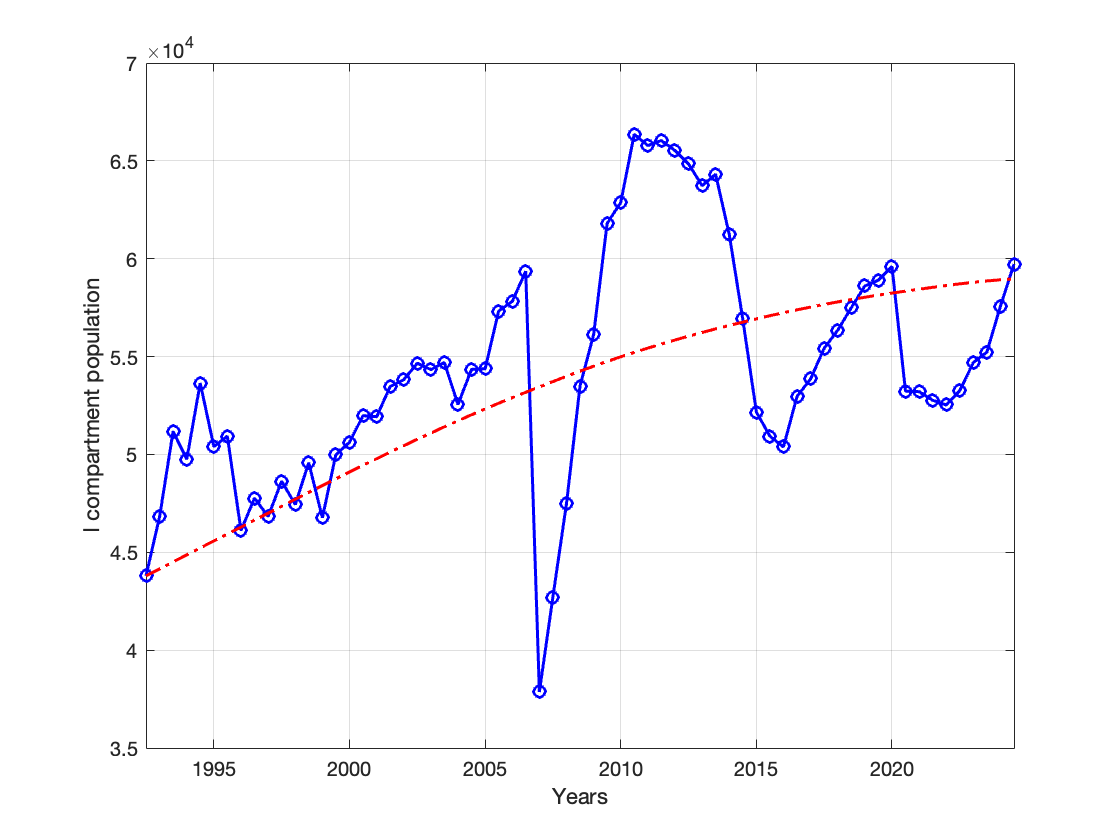}
	\end{center}
	\caption{Observed versus estimated dynamic of the population in the compartment $I$.}
	\label{Fig:obsVSestim}
\end{figure}

The application of the model to real-world data necessarily involves a parameter estimation phase, that is a methodology for determining the values of a model's parameters using empirical data. While certain parameters can be directly measured, others must be inferred through fitting procedures. In practice, parameter estimation assumes that the parameters governing the dynamics of a system are unknown, but that time series data for system inputs and outputs are available. 
As a case study, we applied our model to data provided by the Italian Ministry of Justice\footnote{https://www.giustizia.it/giustizia/page/it/statistiche}. Specifically, we analyzed a time series that records the number of incarcerated individuals and the number of inmates who successfully completed vocational training programs. The dataset consists of semiannual observations spanning from the first half of 1992 to the first half of 2024. These two time series constitute the observables for compartments $I$ and $E$. More precisely, inmates completing a program ''feed'' compartment $E$, so that the observed value $E(t_k)$ at a certain time is given by their cumulative sum.  
For the susceptible population $X$, we considered the adult population in Italy, which was estimated at approximately $N_0 =47,000,000$ in 1992 (source: ISTAT). The model's time step was set to one semester, i.e., $\Delta=0.5$. Several parameters were assigned based on prior demographic and institutional knowledge. For instance, the average detention period in Italy is estimated at approximately $18$ months (Space I 2023 Report, Council of Europe Annual Penal Statistics https://wp.unil.ch/space/space-i/annual-reports/), leading to the choice $\gamma_E =\gamma_I =1/3$ . Additionally, the mortality rate was set at $\mu=0.012$, corresponding to an average life expectancy of approximately $85$ years (see Istat).
Consequently, the parameters to be estimated are $\alpha$ and $\rho$. For this purpose, we focused on the second equation of the dynamic model, following the approach in \cite{CM}. When discretized using the forward Euler method, this yields the following discrete-time dynamical system:
\begin{equation} \label{discrI}
I_{k+1} = I_{k} + \alpha I_k (N_0-I_{k}-E_{k})- \rho I_k E_k - \gamma_I I_k - \mu I_k, \ \ \ k=1, \ldots, M
\end{equation}
having set $X_{k} = N_0-I_{k}-E_{k}$. This dynamical system can be reformulated in a linear regression framework, allowing the application of the classical least squares method to estimate the parameters $\alpha$ and $\rho$. In fact, by setting $\tilde I_{k+1} \equiv I_{k+1} - I_{k}(1-\gamma_I-\mu)$ and $\Phi_k = (\begin{matrix}I_k(N_0-I_k-E_k), & -I_k E_k) \end{matrix}$, we can write (\ref{discrI}) in matrix form as  
$$
\Phi_k \vartheta = \tilde I_{k+1},
$$
where $\vartheta \equiv \left(\begin{matrix} \alpha \\ \rho \end{matrix} \right)$. Given the observed values of the compartments, the standard constrained regression OLS can therefore be used as the basic estimation procedure for the parameters of the model, that is
$$
\hat \vartheta = \mbox{argmin}_{\vartheta \geq 0} \sum_{k=1}^{T-1} || \tilde I_{k+1} - \Phi_k \vartheta ||_2^2,
$$
(see e.g. \cite{CNP20}). The optimal parameter values obtained using such a procedure are 
\begin{eqnarray} \label{param_est}
\hat \alpha & = & 7.4037\cdot 10^{-9}\ \ (0.0069 \cdot 10^{-6}, 0.0079 \cdot 10^{-6}), \\
 \hat \rho & = & 7.7230\cdot 10^{-8} \ \ ( -0.3347 \cdot 10^{-6},  0.4892 \cdot 10^{-6}), 
\end{eqnarray} 
in parenthesis the $95\%$ confidence interval. The corresponding estimated value of the basic reproduction number is $\hat R_0 = 1.0252$, while the threshold is estimated to be $\hat C = 1.0959$. In accordance with Theorem \ref{mainTheorem}, these parameter estimates indicate a regime associated with the education-free equilibrium. The estimated trajectory of $I$, obtained by recursively applying equation (\ref{discrI}) with the estimated parameter values, is compared with the observed data in Figure \ref{Fig:obsVSestim}.

It is important to interpret the results of the fitting exercise with caution, as the estimated values of $R_0$  and the threshold $\overline C$ vary across the extremes of the confidence intervals for $\alpha$ and $\rho$, potentially leading to different qualitative interpretations. In particular, the range of values for $R_0$ is $(0.9617,1.0887)$. In fact, it is well established that estimating $R_0$ is a particularly challenging task (see, e.g. \cite{dslyj19}); a thorough investigation of this issue is beyond the scope of the present study.
Additionally, the dataset used in this analysis does not include information on formal education programs (from elementary school to university level), as such data from the Italian Ministry of Justice are only available starting from 2019. Due to the limited number of observations, the estimation procedure of our model using these data does not yield statistically meaningful results.

\section{Conclusions}

This paper introduces a mathematical model that describes the temporal evolution of crime, treating it as a social epidemic. The model consists of three ordinary differential equations that regulate the movement of individuals between three distinct groups: people who are not incarcerated, offenders who are in prison, and the imprisoned ones who attend an educational program. The first group is considered susceptible, as individuals may become offenders through their social interactions. This perspective on crime as a socially transmitted phenomenon has been explored in various mathematical studies as cited in the Introduction.

The analysis of the dynamical system identifies three equilibrium states: a delinquence free equilibrium, an education free equilibrium (where no inmates pursue educational programs), and a coexistence equilibrium.
The basic reproduction number $R_0$ is derived, and stability analysis using the Jacobian method shows that the delinquence free equilibrium is locally stable when $R_0<1$. Conversely, when $R_0>1$ but remains below the critical threshold $\overline{C}= 1+ \frac{\alpha(\gamma_E+\mu)}{\rho (\gamma_I+\mu)}$, the education free equilibrium becomes stable. Finally if $R_0$ exceeds this threshold, then the coexistence equilibrium point is stable.


Numerical simulations reveal the dynamical properties of the model, in agreement with the theoretical results. Starting from a baseline parameter configuration, we show that varying a single parameter allows the emergence of the three distinct equilibrium states. The simulations highlight the pivotal roles of the parameters $\alpha$ (interaction rate) and $\gamma_I$  (inverse mean detection time) in determining the system's qualitative behavior.
We further applied the model to empirical data provided by the Italian Ministry of Justice, comprising semiannual records from 1992 to 2024 of the total prison population (compartment $I$) and the subset who successfully completed educational programs (compartment $E$). Fixing the demographic and directly interpretable parameters (primarily the average detection time), we fitted the discretized version of the model using linear regression techniques. This allowed estimation of the remaining parameters and, subsequently, the basic reproduction number $R_0$  and the associated threshold $\overline{C}$. The results indicate that the system resides in a regime corresponding to the education-free equilibrium. However, as a final observation, we note that the inherent randomness in the empirical data introduces considerable variability in the parameter estimates, which, in turn, can significantly alter the predicted regime of the system. This underscores the importance of applying advanced time series techniques to filter out observational noise and enhance the reliability of model calibration, like e.g. in \cite{CRS25}. Moreover, expanding the dataset to include additional information, such as participation in various levels of instructional programs (e.g., primary, secondary, or university education), could provide a more comprehensive basis for analysis. Future work may also consider refinements to the model structure to better capture the complexity of the underlying dynamics.

As a matter of fact, several extensions of the model could be considered, both from a theoretical standpoint and an empirical perspective. 
It could be analyzed the reason why it should be desirable to have CE as an equilibrium, introducing some utility function which should consider
detention costs, and social costs together with incentives to participate in educational programs. Finally,  a control problem could be formulated, where the control would act on the number of prisoners involved in education.
Moreover, the inclusion of additional compartments could certainly refine the analysis, although it would introduce non-trivial mathematical complexities. We could consider an extended version of the model with four compartments, essentially dividing the susceptible population into two subgroups: individuals who are susceptible and those who engage in delinquent behavior but do not enter the prison system. 
Further research will be devoted to the study of this extended framework.


\vspace{1cm}

\textbf{Author Contributions.} Both authors equally contributed  in every aspect of the \indent writing of this article. All authors have read and agreed to the published version of 
\indent the manuscript.

\medskip



\textbf{Conflicts of Interests.} The authors declare no conflict of interest.


\begin{thebibliography}{99}
	
	\bibitem{SHIV} Srivastav AK, Ghosh M, Chandra P. Modeling dynamics of the spread of crime in a
	society. Stochastic Anal. Appl., 2019.
	
\bibitem{Pa}	Gonz\'alez-Parra G, Chen-Charpentier B, Kojouharov HV. Mathematical
modeling of crime as a social epidemic. J Interdisc Math,
doi: 10.1080/09720502.2015.1132574, 2018.
	
	\bibitem{BL}	 Blumstein A, Larson RC  Problems in modeling and measuring
	recidivism. Journal of Research in Crime and Delinquency 8: 124--132, 1971.
	
	\bibitem{BBG} Belkin J, Blumstein A, Glass W  Recidivism as a feedback process: An
	analytical model and empirical validation. Journal of Criminal Justice 1: 7--26, 1973.
	
	\bibitem{BLU}		 Blumstein A  An OR missionary's visits to the criminal justice system,	Operations Research 55: 14--23, 2007.
	
	\bibitem{BCR}	 Blumstein A, Cohen J, Roth J, Visher CA  Criminal Careers and Career
	Criminals. National Academy Press, 1986.
	
\bibitem{BLUC}	 Blumstein A, Cohen J,Characterizing criminal careers. Science 238: 985--991, 1982.
	
\bibitem{GAa}	 Greenwood PW, Abrahamse A  Selective incarceration. Rand Report R-
	2815-NIJ, 1982.
\bibitem{CCB}	 Canela-Cacho JA, Blumstein A, Cohen J  Relationship between the
	offending frequency ($\lambda$) of imprisoned and free offenders. Criminology 35: 133--175, 1997.
	
		\bibitem{KM} Kermack, W.O., McKendrick, A.G. Contributions to the mathematical theory of epidemics. \textit{Bulletin of Mathematical Biology} 53 (1--2), 33--55, 1991.
	
	\bibitem{BF2020} Bertozzi, A. L., Franco, E., Mohler, G., Short, M. B., Sledge, D. The challenges of modeling and forecasting the spread of COVID-19. Proceedings of the National Academy of Sciences, 117(29), 16732--16738, 2020.
	
	\bibitem{CRS25} Cerqueti, R., Ramponi, A., Scarlatti, S. A compartmental model for the dynamic simulation of pandemics with a multi-phase vaccination and its application to Italian COVID-19 data, Mathematics and Computers in Simulation, Volume 228, 124--146, ISSN 0378-4754, https://doi.org/10.1016/j.matcom.2024.08.011, 2025.

\bibitem{CPW22} Chen, K., Pun, C. S., Wong, H. Y. Efficient social distancing during the COVID-19 pandemic: Integrating economic and public health considerations. \textit{European Journal of Operational Research}, 304 (1), 84--98, 2022.

	\bibitem{RT23} Ramponi A. and Tessitore M.E. (2023), The economic cost of social distancing during a pandemic: an optimal control approach in the SVIR model, Decision in Economics and Finance, 2023.
	
	
		\bibitem{G}
		M. Gladwell, The Tipping Point: How Little Things Can Make A Big Difference, Little Brown and Company, Boston, 2000.
		
		\bibitem{GAA}
		F. Gino, Ayal S, Ariely D, Contagion and differentiation in unethical behavior, the effect of one bad apple on the barrel, Psychol. Sci. 20, 393--398, 2009.
		
		\bibitem{LT}	 A .A . Lacey, M.N. Tsardakas, A mathematical model of serious and minor criminal activity, Eur. J. Appl. Math. 27, 403--421, 2016.
		
		\bibitem{So} J. Sooknanan, D.M.G. Comissiong, When behaviour turns contagious: the use of deterministic epidemiological models in modeling social contagion
		phenomena, Int. J. Dyn. Contr. 5  1046--1050, 2017.
		
		\bibitem{DP} M.R. D'Orsogna, M. Perc, Statistical physics of crime: a review, Phys. Life Rev. 1,2  1--21,  2015.
		
		
		\bibitem{SB}	 J. Sooknanan, B. Bhatt, D.M.G. Comissiong, A modified predator-prey model for the interaction of police and gangs, R. Soc. Open Sci. 3, 2016.
		
		\bibitem{GSM}		A. Goyal, J. Shukla, A. Misra, and A. Shukla. Modeling the role of
		government efforts in controlling extremism in a society. Mathematical
		Methods in the Applied Sciences, 2014.
		
		\bibitem{McM}	 D. McMillon, C. P. Simon, and J. Morenoff. Modeling the underlying
		dynamics of the spread of crime. PLoS ONE, 9(4):e88923, 04 2014.
		
		\bibitem{Mis}	 A. Misra. Modeling the effect of police deterrence on the prevalence
		of crime in the society. Applied Mathematics and Computation, 237:531--545, 2014.
		
		\bibitem{MBB}	 M. B. Short, P. J. Brantingham, and M. R. D'Orsogna. Cooperation
		and punishment in an adversarial game: How defectors pave the way
		to a peaceful society. Phys. Rev. E, 82:066114, 2010.
		
		\bibitem{CC}	R. Cantrell, C. Cosner, and R. Manasevich. Global bifurcation of solutions
		for crime modeling equations. SIAM Journal on Mathematical
		Analysis, 44(3):1340?1358, 2012.
		
		\bibitem{MPS}	R. Manasevich, Q. H. Phan, and P. Souplet. Global existence of solutions
		for a chemotaxis-type system arising in crime modelling. European
		Journal of Applied Mathematics, 24:273?296, 4, 2013.
		
		\bibitem{MS}		G. Mohler and M. Short. Geographic profiling from kinetic models
		of criminal behavior. SIAM Journal on Applied Mathematics, 72(1):163--180, 2012.
		
		\bibitem{ABP}	 A. B. Pitcher. Adding police to a mathematical model of burglary.
		European Journal of Applied Mathematics, 21(4-5):401--419, 2010.
		
		\bibitem{RB}	 N. Rodriguez and A. Bertozzi. Local existence and uniqueness of solutions
		to a pde model for criminal behavior. Mathematical Models and
		Methods in Applied Sciences, 20(supp01):1425--1457, 2010.
		
		\bibitem{SBB}	 M. Short, A. Bertozzi, and P. Brantingham. Nonlinear patterns in urban
		crime: Hotspots, bifurcations, and suppression. SIAM Journal on
		Applied Dynamical Systems, 9(2):462--483, 2010.
		
		\bibitem{SDO}	 M. B. Short, M. R. D'Orsogna, V. B. Pasour, G. E. Tita, P. J. Brantingham,
		A. L. Bertozzi, and L. B. Chayes. A statistical model of criminal
		behavior. Mathematical Models and Methods in Applied Sciences,
		18(supp01):1249--1267, 2008.
		
		\bibitem{SGA}	 J. B. Shukla, A. Goyal, K. Agrawal, H. Kuswah, and A. Shukla. Role of
		technology in combating social crimes: A modeling study. European
		Journal of Applied Mathematics, 24:501--514, 8 2013.
		
		\bibitem{Tao}	 W. Tao. Computational criminology and evolution mechanisms of social
		crime dynamic system. In Electronics, Computer and Applications,
		2014 IEEE Workshop on, pages 481--484, May 2014.
		
		\bibitem{YT}	 T. Yokoyama and T. Takahashi. Mathematical neurolaw of crime and
		punishment: The q-exponential punishment function. Applied Mathematics,
		4, 1371--1375, 2013.
		
		\bibitem{BC} Brauer, F. and Castillo-Chavez, C. Mathematical Models in Population Biology and Epidemiology. Springer Science, Berlin, 2010.
		
		\bibitem{KDM}T. Kwofie, M. Dogbatsey, S.E. Moore	Curtailing Crime Dynamics: A Mathematical Approach, Frontiers in Applied Mathematics and Statistics, 2023.
		
			\bibitem{SC} Sooknanan J., Comissiong  D.M.G., A mathematical model for the treatment of delinquent behaviour, Socio--Economic Planning Sciences,  v.63, pp. 60--69, 2018.
			
		\bibitem{CM} Calatayud, J., Jornet, M., \& Mateu, J.  A dynamical mathematical model for crime evolution based on a compartmental system with interactions, International Journal of Computer Mathematics, 102(1), 44--59, 2024.
		
		\bibitem{CNP20} Calafiore, G.C., Novara, C., Possieri, C., A time-varying SIRD model for the COVID-19 contagion in Italy, Annual Reviews in Control, 50, 361--372, 2020.
		
		\bibitem{dslyj19} Delamater PL, Street EJ, Leslie TF, Yang YT, Jacobsen KH. Complexity of the Basic Reproduction Number (R0). Emerg Infect Dis.,25(1):1-4. doi: 10.3201/eid2501.171901. PMID: 30560777; PMCID: PMC6302597, 2019.
\end{thebibliography}
\end{document}